\documentclass[sigconf]{acmart}
\AtBeginDocument{%
  }

\setcopyright{acmlicensed}

\copyrightyear{2026}
\acmYear{2026}
\setcopyright{cc}
\setcctype{by-nc-nd}
\acmConference[SAC '26]{The 41st ACM/SIGAPP Symposium on Applied Computing}{March 23--27, 2026}{Thessaloniki, Greece}
\acmBooktitle{The 41st ACM/SIGAPP Symposium on Applied Computing (SAC '26), March 23--27, 2026, Thessaloniki, Greece}
\acmPrice{}
\acmDOI{10.1145/3748522.3779807}
\acmISBN{979-8-4007-2294-3/2026/03}

\usepackage{algorithm}
\usepackage{algpseudocode}
\usepackage{graphicx}
\usepackage{textcomp}
\usepackage{xcolor}
\usepackage{multirow}
\usepackage{threeparttable}
\usepackage{booktabs}
\usepackage{colortbl}
\definecolor{lightgray}{gray}{0.9}
\usepackage{array}  %

\newcolumntype{C}[1]{>{\centering\arraybackslash}p{#1}}  %

\begin{document}

\title{DEFEND: Poisoned Model Detection and Malicious Client Exclusion Mechanism for Secure Federated Learning-based \\ Road Condition Classification}
\renewcommand{\shorttitle}{DEFEND: Detection and Exclusion Mechanism for Secure FL-RCC}


\author{Sheng Liu}
\affiliation{%
  \institution{Networked Systems Security Group \\ KTH Royal Institute of Technology}
  \city{Stockholm}
  \country{Sweden}}
\email{shengliu@kth.se}

\author{Panos Papadimitratos}
\affiliation{%
  \institution{Networked Systems Security Group \\ KTH Royal Institute of Technology}
  \city{Stockholm}
  \country{Sweden}}
\email{papadim@kth.se}



\renewcommand{\shortauthors}{S. Liu and P. Papadimitratos}

\begin{abstract}
Federated Learning (FL) has drawn the attention of the Intelligent Transportation Systems (ITS) community. FL can train various models for ITS tasks, notably camera-based Road Condition Classification (RCC), in a privacy-preserving collaborative way. However, opening up to collaboration also opens FL-based RCC systems to adversaries, i.e., misbehaving participants that can launch Targeted Label-Flipping Attacks (TLFAs) and threaten transportation safety. Adversaries mounting TLFAs poison training data to misguide model predictions, from an actual source class (e.g., wet road) to a wrongly perceived target class (e.g., dry road). Existing countermeasures against poisoning attacks cannot maintain model performance under TLFAs close to the performance level in attack-free scenarios, because they lack specific model misbehavior detection for TLFAs and neglect client exclusion after the detection. To close this research gap, we propose DEFEND, which includes a poisoned model detection strategy that leverages neuron-wise magnitude analysis for attack goal identification and Gaussian Mixture Model (GMM)-based clustering. DEFEND discards poisoned model contributions in each round and adapts accordingly client ratings, eventually excluding malicious clients. Extensive evaluation involving various FL-RCC models and tasks shows that DEFEND can thwart TLFAs and outperform seven baseline countermeasures, with at least 15.78\% improvement, with DEFEND remarkably achieving under attack the same performance as in attack-free scenarios.  
\end{abstract}



\maketitle

\section{Introduction}\label{sec_intro}

The automated identification of road surface conditions, e.g., unevenness level, friction magnitude, and material type, is an important tool for Intelligent Transportation Systems (ITS) \cite{CHEN2025125978}. For example, when an autonomous vehicle detects in real-time a waterlogged road ahead through its cameras and \textit{Road Condition Classification (RCC)} model (notably Deep Neural Network, DNN), it can proactively adjust its speed, activate the traction control system, or reconfigure the suspension. Capitalizing on \textit{image data} is particularly feasible for RCC \cite{8569396} due to the widespread availability and cost-effectiveness of on-board cameras, capturing rich contextual details essential for classification accuracy. 

If each vehicle relied only on its own data for model training, RCC would suffer from an inherently skewed perception; e.g., a vehicle driving mostly in the countryside would lack data and the ability to classify inputs from a city center environment. The \textit{centralized} data collection, with all data uploaded to a server to perform model training is impractical too in the long term: 1) regulations about user privacy\footnote{https://gdpr-info.eu/}$^,$\footnote{https://www.oag.ca.gov/privacy/ccpa}$^,$\footnote{http://en.npc.gov.cn.cdurl.cn/2021-12/29/c\_694559.htm} are increasingly strict, 2) computation and energy costs at the data center leave vehicle resources underutilized; and 3) transmitting original data leads to high bandwidth usage and increased response latency for vehicles. \textit{Federated Learning (FL)} \cite{mcmahan2017communication,AFM3D,you2023federated}, notably cross-device horizontal FL, provides a promising solution for the above-mentioned dilemma between regulation considerations and resource utilization, while collecting user contributions across a large-scale deployment with varying environments. Through an iterative client-server collaboration on model parameters, a top-performing RCC model can be learned \cite{10606293}.

\begin{figure}[t]
\centerline{\includegraphics[width=0.473\textwidth]{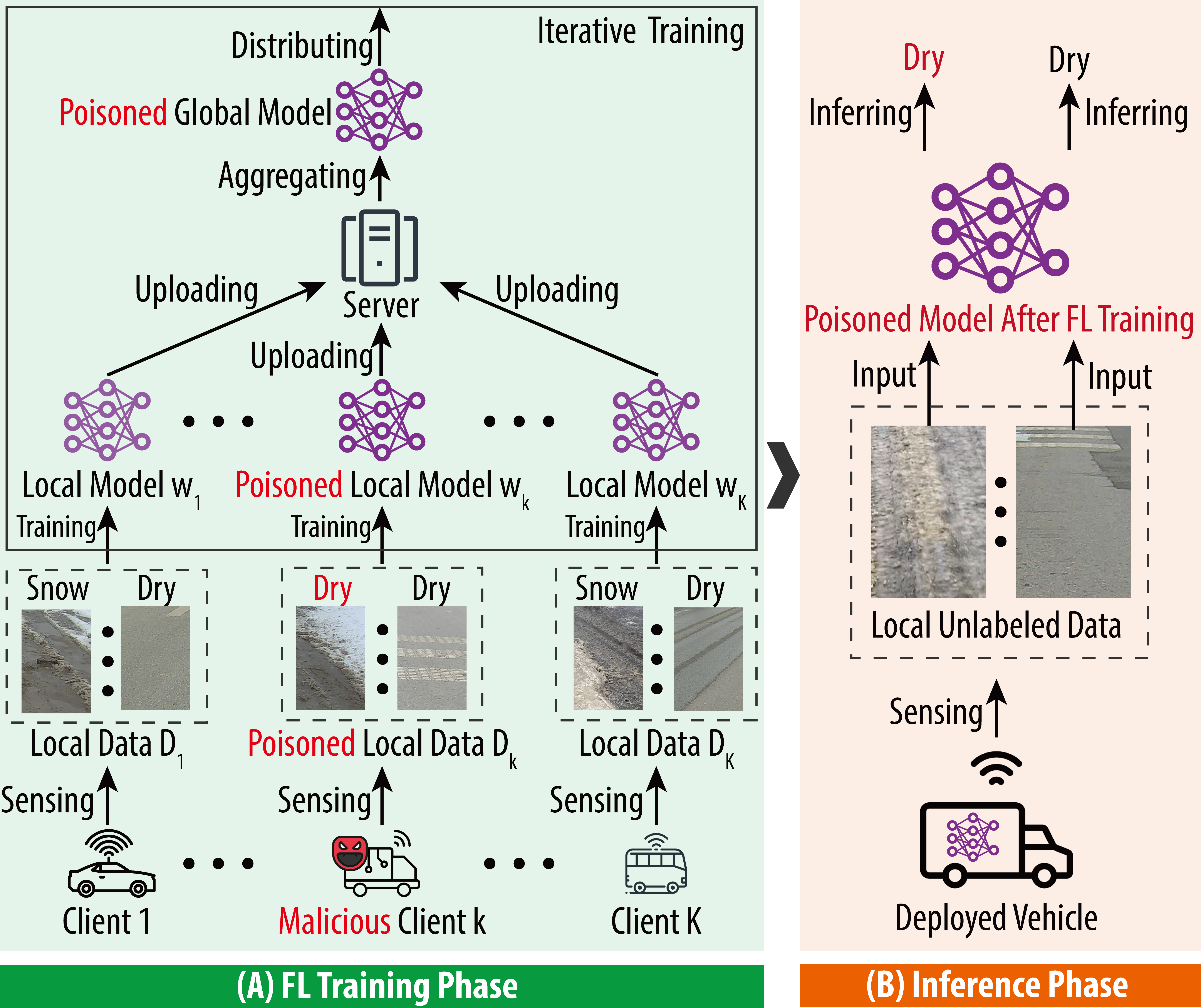}}
\vspace{-1.0em}
\caption{Illustration of TLFAs in FL-RCC. (A) Training Phase: Adversaries deliberately mislabel their data; their local models are poisoned after local training, and the global model is poisoned after global aggregation. (B) Inference Phase: Vehicles equipped with the learned model would predict wrong road conditions that threaten transportation safety.}
\vspace{-1.5em}
\Description{}
\label{fig_TLFA_RCC}
\end{figure}

However, FL-based RCC systems relying on potentially any participant are vulnerable to compromised, adversarial clients. Even though credential management, access control, and secure communication \cite{4689252} can be a first line of strong defense, the vulnerability to \textit{Targeted Label-Flipping Attacks (TLFAs)} \cite{sameera2024lfgurad} remains: malicious clients, i.e., internal adversaries, deliberately change their image labels from a source class (true class) to a target class (falsified class), and use such poisoned data for local model training, thus misleading the prediction results of the aggregated global model. For example, as illustrated in Figure \ref{fig_TLFA_RCC}, during the FL training phase, if the adversary flips road friction labels from “snow” to “dry”, its local model would learn wrongly based on mislabeled data, thus the global model would also be poisoned after each global aggregation round in which such an adversary participates. During the inference phase, the consequent RCC model may identify actual snow conditions as dry, an underestimation of hazardous road conditions jeopardizes transportation safety and may increase accident rates. Results in \cite{LiuP:C:2025c} and in Section \ref{sec_result_anal} both indicate huge performance degradation for FL-RCC when TLFAs happen without defense. Moreover, compared to other more complicated attacks, such as backdoor attacks \cite{li2024backdoorindicator}, TLFAs only require simpler replacement operations on labels (not replacement or manipulation of original image pixels or features), thus they are practical from an adversarial point of view, in the sense that they are efficient to launch with vehicle resources.

Existing defenses mitigating poisoning attacks face two primary challenges in the context of FL-RCC. First, general \textit{poisoned model detection} methods, such as FoolsGold \cite{fung2020limitations} and FLAME \cite{nguyen2022flame}, struggle with the Non-Independent and Identically Distributed (Non-IID) nature of vehicular data. Vehicle driving behaviors vary across locations and over time, yielding markedly different client distributions. General detection methods \cite{fung2020limitations,nguyen2022flame} that are not tailored to TLFAs can fail because poisoned models can still easily appear as benign in a heterogeneous environment, as legitimate updates themselves may exhibit large variance. Recent novel poisoning attack mitigation methods primarily focus on backdoor attacks \cite{fereidooni2023freqfed} or untargeted attacks \cite{10.1145/3488932.3517395}, thus they are not specifically designed for TLFAs. On the other hand, current countermeasures pay attention to model-level misbehavior detection, while missing an effective joint vehicle-level \textit{malicious client exclusion} strategy based on model-level detection results. By uploading poisoned models, malicious clients can consistently threaten the FL-RCC system if they are not excluded. The state-of-the-art countermeasure against TLFAs for FL-RCC, FLARE \cite{LiuP:C:2025c}, still has difficulties in reducing the mis-classification level to match the performance in an attack-free scenario when such attacks happen.

To close this gap, this paper proposes DEFEND: poisoned model \underline{D}etection and malicious client \underline{E}xclusion mechanism for \underline{FE}derated learning-based road co\underline{ND}ition classification. DEFEND first identifies source and target classes (neurons) in TLFAs based on neuron-wise magnitude analysis of the DNN's output layer. With the two recognized neurons, two strategies are implemented in each round: 1) model parameters directly connected to source and target neurons are extracted and clustered by the Gaussian Mixture Model (GMM) to detect poisoned model contributions, which are then filtered out before aggregation; 2) two core metrics evaluating TLFAs, Source Recall (SRec)\footnote{The fraction of source class samples that are correctly classified.} and Attack Success Rate (ASR)\footnote{The ratio of samples with the source label misclassified into the target class.}, are calculated to monitor the global model performance; if these metric values achieve predefined performance change thresholds, the current global model is still considered poisoned even after the local model filter and is discarded. Finally, leveraging the aforementioned poisoned-model detection, DEFEND introduces an adaptive client rating strategy based on decision theory \cite{PapadimitratosH:J:2006}. A client rating is decremented if its contribution is deemed poisoned in a round and incremented otherwise. Once its rating falls below a threshold, the client is identified as malicious with high confidence and it is promptly excluded from further participation in the FL training. Extensive experiments across three DNNs (ResNet-18 \cite{he2016deep}, EfficientNet-B1 \cite{tan2019efficientnet}, and Deit-Tiny \cite{touvron2021training}) and three RCC tasks (friction, material, and unevenness classification) show that the proposed DEFEND outperforms seven baselines (FedAvg \cite{mcmahan2017communication}, Krum \cite{blanchard2017machine}, Trimmed Mean (TMean) \cite{yin2018byzantine}, Median \cite{yin2018byzantine}, FoolsGold \cite{fung2020limitations}, FLAME \cite{nguyen2022flame}, and FLARE \cite{LiuP:C:2025c}) in terms of three evaluation metrics (Global Accuracy (GAcc), SRec, and ASR). 

In brief, the main contributions of this paper are:
\begin{enumerate}
    \item An improved poisoned model detection strategy specifically designed for FL-RCC systems under TLFAs, based on neuron-wise magnitude analysis and GMM clustering. 
    \item An enhanced malicious client exclusion strategy based on adaptive client rating using decision theory leveraging the aforementioned model-level misbehavior detection.
    \item Significantly improved defense against TLFAs compared to the state-of-the-art: DEFEND is the first to maintain model performance equal to that in an attack-free scenario, on average outperforming the best baseline defense by 2.81\%, 24.99\%, and 15.78\% regarding GAcc, SRec, and ASR, respectively.
\end{enumerate}

The rest of this paper is organized as follows. Section \ref{sec_rw} reviews camera-based FL-RCC, corresponding TLFAs, and defenses. Section \ref{sec_sm} describes the system and adversary models. Our scheme, DEFEND, is introduced and evaluated in Section \ref{sec_DEF} and Section \ref{sec_eva}, respectively. Finally, Section \ref{sec_con} concludes this paper and outlines future work.

\section{Background and Related Work}\label{sec_rw}


\subsection{Federated Learning (FL)-based Road Condition Classification (RCC)}

Given that road condition changes, for example when the weather changes, can be local and unpredictable, RCC is important for smart vehicles \cite{MALIN2019181} to timely adjust their braking, steering, suspension, and other driving safety and driver assistance systems. RCC can be camera-based \cite{zhao2023comprehensive, 10705359}, as is the focus of this paper, but it can also rely on other sensors \cite{8896015,varona2020deep}. A key distinction is that cameras are and will be increasingly available, allowing car systems to perceive a gamut of conditions with one sensor (the camera), offering data and early detection, e.g., of a water puddle or cracked asphalt before hitting them; at which point, for example, an inertial sensor could have given relevant data but only post-facto.

To leverage privacy-sensitive image data and distributed on-board resources, the feasibility of FL-based RCC systems was recently explored. Specifically, FedRD \cite{YUAN2021385} achieves individual-level privacy protection and high-performing hazardous road damage detection; Then, its follow-up works, FLRSC \cite{10422129} and FedRSC \cite{10606293}, support multi-label RCC tasks in FL. However, current FL-RCC proposals primarily focus on improving the classification performance of models and the privacy guarantees for vehicles. How to secure FL-RCC systems still remains a largely open problem. 

\subsection{Poisoning Attacks against FL-RCC}

TLFAs \cite{tolpegin2020data,3670434} are essentially Data Poisoning Attacks (DPAs) \cite{tolpegin2020data,10054157,khuu2024data}. Model poisoning attacks \cite{shejwalkar2021manipulating} demand higher levels of adversarial expertise and specialized knowledge (e.g., understanding model architectures and training configurations) as well as greater adversarial computational resources (e.g., for manipulating millions of model parameters) \cite{jebreel2024lfighter,sameera2024lfgurad}; thus they are not as practical as DPAs if launched by resource-restricted on-board vehicle platforms. Furthermore, as local models are sent to the server for aggregation, without the server accessing client local data for privacy considerations, DPAs are more difficult to detect compared to model poisoning attacks.

DPAs can be divided into untargeted DPAs and targeted DPAs \cite{sameera2024lfgurad}. The former degrades the overall performance of the global model and thus can be more easily detected and countered by the system before model deployment; the latter targets specific inputs, thus can be easier to disguise, especially in a heterogeneous environment. As some misclassifications in RCC are more dangerous than others, targeted DPAs, notably TLFAs, are more practical and significant than untargeted DPAs in our investigation here. To see why this is the case, let us revisit the above-mentioned example of road friction level: mislabeling the actual “snow” road surface as “dry” may have vehicle safety implications, while the opposite misclassification, “dry”$\rightarrow $“snow”, reduces traffic efficiency.  

\subsection{Defensive Mechanisms for FL-RCC}

Various countermeasures are proposed to defend against general DPAs in FL, without considering neither RCC nor TLFAs. FoolsGold \cite{fung2020limitations} assumes that poisoned models are more similar than benign models; thus, it first calculates the cosine similarity between local model output layers and then penalizes the updates with larger similarity to mitigate potential poisoning attacks. CONTRA \cite{CONTRA} also measures the cosine similarity of local updates to record credibility for promotion or penalty of updates. FLAME \cite{nguyen2022flame} utilizes jointly differential privacy, model clustering, and weight clipping technologies, trading off global model performance for adversary detection efficiency. However, the data in FL-RCC is inherently Non-IID, as distinct spatial and temporal vehicle driving patterns lead to diverse data distributions. The intrinsic heterogeneity complicates the task of distinguishing malicious updates from benign ones \cite{fereidooni2023freqfed}, as innocent updates also have high variance, leading to significant countermeasure performance degradation on RCC. More refined features specific to TLFAs should be extracted, and more effective detection mechanisms should be designed, taking the mentioned heterogeneity into consideration. 

Recent novel countermeasures against poisoning attacks pay attention to mitigating backdoor attacks. FreqFed \cite{fereidooni2023freqfed} capitalizes on the discrete cosine transform to distinguish good and bad updates in the frequency domain. CrowdGuard \cite{rieger2022crowdguard} analyzes hidden layer outputs of local models and executes iterative pruning to detect backdoors. However, such attacks/adversaries elaborately falsify both the original image data and the corresponding labels. These attacks and their corresponding defenses are fundamentally different from TLFAs, adding triggers, and DEFEND, utilizing the whole output layer. 

Only few defensive mechanisms are specifically designed for vehicles. LFGurad \cite{sameera2024lfgurad} first proposes a hierarchical FL framework for vehicular networks, then feeds the activations of the output layer in each local model into a multi-class support vector machine for malicious model classification; finally, it is evaluated on the structured traffic sign classification dataset. RoHFL \cite{10046398} is another robust hierarchical FL framework that develops a logarithm-based normalization method to address maliciously scaled model parameters. OQFL \cite{9641742} uses a quantum-behaved particle swarm optimization method that can automatically update hyper-parameters in FL against adversarial attacks targeted for autonomous driving. Both RoHFL and OQFL are evaluated on general classification datasets such as MNIST and Fashion-MNIST. However, they all focus on passive misbehavior detection at the model level, while ignoring active malicious client exclusion at the vehicle level. 

FLARE \cite{LiuP:C:2025c} is the state-of-the-art solution designed for FL-RCC against TLFAs; however, both its HDBSCAN-based model clustering method and its count-based client filtering method are moderately effective, with a very significant RCC performance gap when comparing FLARE under attack and a TLFA-free situation. Specifically, the SRec value for a no-attack scenario is 80\%; and it drops to 60\% when under TLFAs, with a similar gap for ASR values, 5\% vs. 30\%. In contrast, DEFEND thwarts TLFAs, achieving SRec and ASR of about 80\% and 5\%, respectively, even under attack. All in all, under TLFAs, there is no existing countermeasure that can achieve the same FL-RCC model performance as the TLFA-free scenario.

\section{System Model and Adversary Model}\label{sec_sm}

\subsection{System Model}

\textbf{Vehicular Protocols for Secure and Private Communication:} We consider an FL-RCC system with one trusted server and a massive set of available clients to contribute, each registered via a cloud‐based Vehicular Public Key Infrastructure (VPKI) \cite{10075082,KhodaeiNP:J:2023}. Vehicles obtain short‐lived pseudonym certificates \cite{KhodaeiJP:J:2018}, syntactically unlinkable and anonymized credentials issued for minutes or hours, to guarantee authenticity, integrity, non‑repudiation, and privacy (conditional anonymity and long-term unlinkability). This design choice ensures compatibility with standardized security and privacy for cooperative ITS systems, notably V2X (Vehicle to Vehicle/Infrastructure) \cite{10075082, 4689252, KhodaeiP:J:2015}. Misbehavior attributed to one or more pseudonyms can lead to rapid revocation \cite{KhodaeiP:J:2021b}, thus timely efficient eviction of the wrongdoer from the system. Clients establish end‑to‑end confidential and authenticated channels with the server over TLS \cite{rescorla2018transport}, using their current pseudonym, ensuring secure and privacy-preserving delivery of model updates.

\textbf{FL Procedure:} $K$ clients $\mathbb{C}=\{c_1,c_2,...,c_K\}$ form a cluster to undertake the current RCC model training task. The task involves $E$ classes/labels and a sequence $S=[1,2,...,E]$ to encode them, where larger numbers represent more dangerous road condition classes. Consequently, output layer $L$ consists of $E$ neurons: $L=[l_1,l_2,...,l_E]$. Each client, $c_k$, owns a private dataset, $\mathbb{I_k}$, with $n_k$ image-label pairs, where images are captured by on-board cameras and could be labeled by driver feedback or annotation tools \cite{10480248}. In each FL round $t$, a subset of clients $\mathbb{C}^t\subseteq\mathbb{C}$ are randomly selected for local training, where $|\mathbb{C}^t|=M$, $M\leq K$. After receiving the newest global model $\omega^t$ from the server, each participant $c_k\in \mathbb{C}^t$ updates the model to $\omega^t_k$ using $\mathbb{I}_k$ according to Equation (\ref{eq_sgd}), where $\eta$ represents the learning rate, and ${\mathcal {L}}_{k}$ denotes the loss function of client $c_k$, e.g., cross-entropy. Finally it sends $\omega^t_k$ back to the server for aggregation.

\begin{equation}\label{eq_sgd}
    \omega^t_k=\omega^t-\eta \nabla _{\omega^t}{\mathcal {L}}_{k}(\omega^t,I_{k})
\end{equation}

A new global model $\omega^{t+1}$ is formed as Equation (\ref{eq_agg}), where $q_k$ is the aggregation weight assigned to $c_k$. Generally speaking, $q_k$ is the normalized data size \cite{mcmahan2017communication}; however, to avoid malicious clients providing falsified information that could increase the attack impact, here we set $q_k$ as $\frac{1}{M}$. It is also important to note that this paper focuses on mitigating TLFAs, rather than improving the no-attack model performance. Moreover, countermeasures against TLFAs, including ours, do not rely on specific aggregation strategies. Thus, we choose Equation (\ref{eq_agg}) for security and simplicity.

\begin{equation}\label{eq_agg}
    \sum_{c_k\in \mathbb{C}^t}q_k\times \omega^t_k
\end{equation}

These local training and global aggregation processes are executed iteratively until the global model converges. The final learned model is deployed to automated vehicles, over secure server-client communication, to support improved real-time RCC with unseen road surface images. 

\subsection{Adversary Model}

With $P$ malicious clients in the system, where $P\leq \frac{K}{2}$, each adversary flips its local labels from a source class $f$ to a less hazardous target class $g$ (without altering input features), then trains its local model on the poisoned data. After that, each adversary submits corrupted local updates to the server to poison the global model during the aggregation process. The goal of malicious clients is to selectively degrade the model performance on the source class, thereby causing vehicles to underestimate hazard levels (e.g., misclassifying snow roads as dry, as shown in Figure \ref{fig_TLFA_RCC}); which poses a greater safety risk than uniform performance degradation (e.g., untargeted poisoning attacks randomly flip labels without a specific misclassification goal). 

To do so, $f$ should be smaller than $g$ (if $f>g$, the attack goal is disrupting traffic efficiency rather than safety, e.g., flipping from actual dry roads to snow). Strategies of the adversary can also be adaptive, i.e., choosing different source and/or target classes during specific rounds, aiming to bypass potential defensive mechanisms. 

We assume that adversaries cannot compromise the trusted server, and they cannot control the random client selection process (choosing $\mathbb{C}^t$) at the server side. We do not dwell on the introduction of malicious clients in the system - they can be gradually registered with the system, provisioned with credentials, and modified functionality that deviates from the system specification (or similarly, they can be compromised clients, e.g., with the FL functionality adversarially modified).  

\section{Our Scheme: DEFEND}\label{sec_DEF}

\textbf{Scheme Overview:} The poisoned model \underline{D}etection and malicious client \underline{E}xclusion mechanism for \underline{FE}derated learning-based road co\underline{ND}ition classification (DEFEND) workflow in each round is illustrated in Figure \ref{fig_DEFEND}. Algorithm \ref{alg_1}, \textit{Line 1} initializes a client blacklist $\mathbb{B}$, an SRec value, an ASR value, and for each client $c_k$ a rating value $r_k(0)$. \textit{Lines 2-8} randomly select clients not in the blacklist to execute local model training and upload updates in each round, protected by security and privacy protocols, notably pseudonymously authenticated TLS. \textit{Lines 9-15} analyze neuron-wise magnitudes regarding output layer $L$ to identify source and target classes as $f'$ and $g'$. \textit{Lines 16-17} detect poisoned models via a Gaussian Mixture Model (GMM) based on $U^t$: value changes of parameters connected to $f'$ and $g'$. \textit{Lines 18-24} validate the new global model, $\omega^{t+1}$, based on SRec and ASR values to decide accept or discard it. \textit{Lines 25-34} update rating values and the blacklist to exclude malicious clients. Each of the detected outliers in $\mathbb{C}^t_{out}$ sends model parameters over each of the secure channels in a non-repudiable manner. Given the use of a valid pseudonym (contributions of a given client are anonymized yet they can be linked to each other across FL rounds), client rating can be reduced, so that a deemed malicious client can be excluded from current and future FL processes, while at the same time rendering a client participation in different FL executions unlinkable. 

\begin{figure}[tb]
\centerline{\includegraphics[width=0.48\textwidth]{figures/DEFEND-Workflow.png}}
\vspace{-1.0em}
\caption{Workflow of DEFEND in round $t$. Steps (1) and (3) extract features and are executed in parallel for each local model. Step (2) identifies source and target neurons in TLFAs. Steps (4)-(6) execute the local model filtering, malicious client exclusion, and global model validation, respectively. Note that output layer parameters are marked in blue, while source and target neuron parameters are marked in red.}
\vspace{-1.0em}
\Description{}
\label{fig_DEFEND}
\end{figure}

\begin{algorithm}[tb]
\caption{Protocol of DEFEND}
\label{alg_1}
\begin{algorithmic}[1]
\State Initialize black list $\mathbb{B} = \emptyset$, $SRec^{old}=0$, $ASR^{old}=1$, and rating value $r_k(0)=\delta(r^{max}-r^{min})$ for each client $v_k$
\For{each round $t \in [1,T]$}
    \State $\mathbb{C}^t \gets$ randomly select $M$ clients from $\mathbb{C}-\mathbb{B}$
    \State The server sends $\omega^t$ to all clients in $\mathbb{C}^t$
    \For{each client $c_k \in \mathbb{C}^t$ in parallel}
        \State Update local model $\omega^t_k$
        \State Send $\omega^t_k$ back to the server
    \EndFor
    \State The server receives $\omega^t_k$ from $\mathbb{C}^t$
    \For{each $\omega^t_k$ the server}
        \State $\Delta^t_{k,L}=\{\omega^t_{k,l}-\omega^t_l|l\in L\}$ \Comment{Output layer changes}
        \State Calculate magnitudes $||\Delta^t_{k,l}||_2$ for $l\in L$ \Comment{$\ell_2$-norm}
    \EndFor
    \State $\{||\Delta^t_{l_{1}}||_2,...,||\Delta^t_{l_{E}}||_2\}\gets$ neuron-wise magnitudes
    \State $f', g'$ $\gets$ Top-2($\{||\Delta^t_{l_{1}}||_2,...,||\Delta^t_{l_{E}}||_2\}$)\Comment{$f'<g'$}
    \State $U^t \gets \{\Delta^t_{k,l}|c_k \in \mathbb{C}^t, l \in \{g', f'\}\}$
    \State $\mathbb{C}^t_{out}$ = GMM($U^t$)
    \State $\omega^{t+1}=Aggeregate\{\omega^t_{k}|c_k \notin \mathbb{C}^t_{out}\}$
    \State $SRec^{new}, ASR^{new}$ $\gets$ Validate ($\omega^{t+1}$)
    \State $\Delta SRec=SRec^{new}-SRec^{old}$, $\Delta ASR=ASR^{new}-ASR^{old}$
    \If{$\Delta SRec<SRec^{thr}$ or $\Delta ASR>ASR^{thr}$}
        \State $\omega^{t+1}=\omega^{t}$
    \EndIf
    \State $SRec^{old}=SRec^{new}$, $ASR^{old}_k=ASR^{new}_k$
    \For{$c_k \in \mathbb{C}^t$}
        \If{$c_k \in \mathbb{C}^t_{out}$}
            \State $r_k(t)=max\{r_k(t-1)-\gamma, r_{min}\}$
            \If{$r_k(t)\leq r^{min}$ $And$ $c_k \notin \mathbb{B}$ }
                \State Add $c_k$ in $\mathbb{B}$ 
            \EndIf
        \Else
            \State $r_k(t)=min\{r_k(t-1)+\beta, r^{max}\}$
        \EndIf
    \EndFor
\EndFor
\State \Return $\omega^{T+1}$
\end{algorithmic}
\end{algorithm}

\textbf{Poisoned Model Detection:} Poisoned models trained under TLFAs and honest models trained normally have contradictory objectives \cite{jebreel2024lfighter}, resulting in more significant differences for those parameters directly connected to source and target neurons. Thus, in each round, after getting the value changes regarding output layer as $\Delta^t_{k,L}=\{\omega^t_{k,l}-\omega^t_l|l\in L\}$ for each $c_k$, we calculate the magnitude ($\ell_2$-norm) for each output neuron $l$ as $\lVert\Delta^t_{k,l}\rVert_2$ according to Equation (\ref{eq_client-neuron-norm}), where $d_l$ is the number of parameters associated with output neuron $l$.

\begin{figure}[tb]
\centerline{\includegraphics[width=0.48\textwidth]{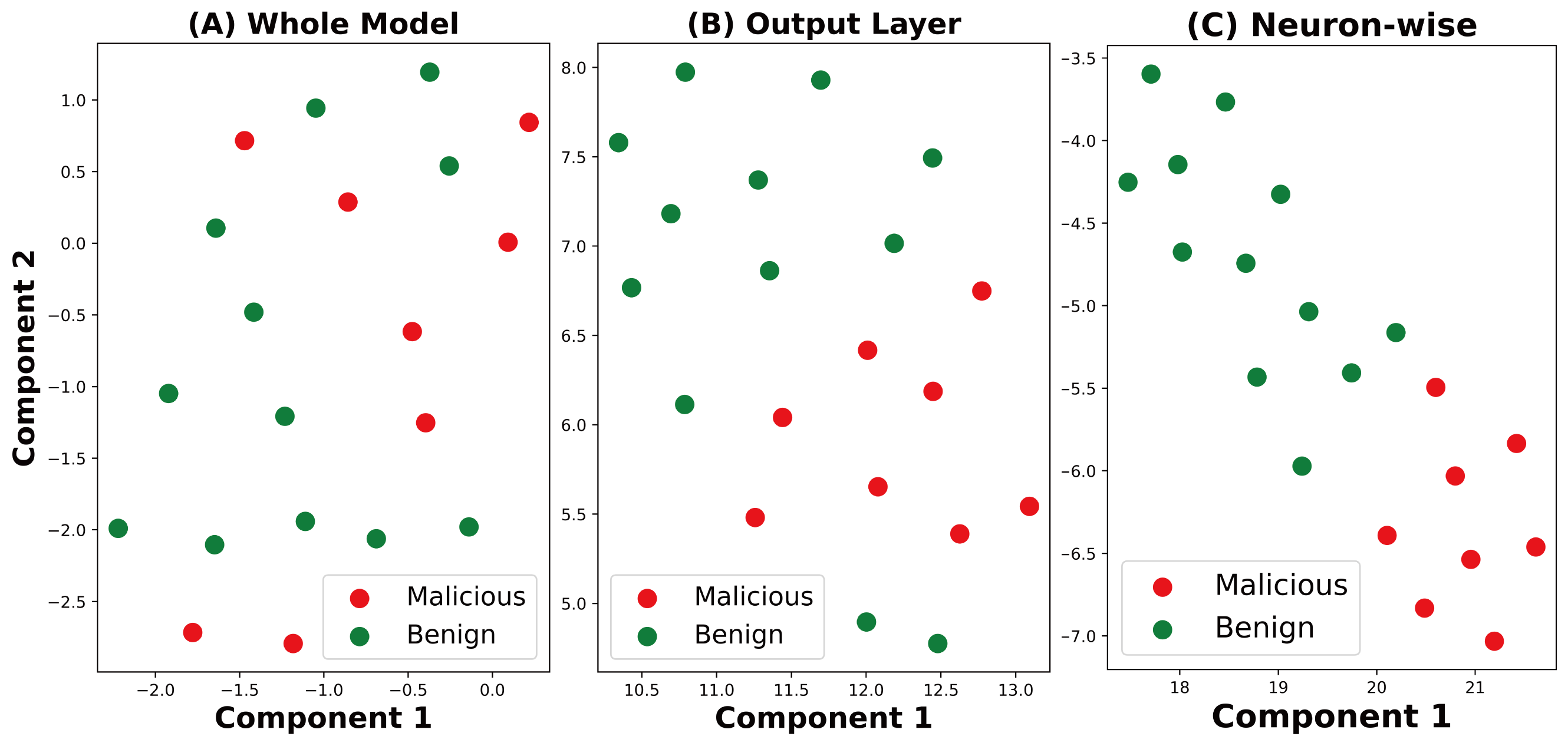}}
\vspace{-1.0em}
\caption{Comparison between malicious and benign updates based on three kinds of features: (A) whole model parameters, (B) output layer parameters, and (C) neuron-wise parameters (with two more distinctive clusters).}
\vspace{-1.0em}
\Description{}
\label{fig_scatter}
\end{figure}

\begin{equation}
\lVert \Delta^t_{k,l}\rVert_2
= \sqrt{\sum_{i=1}^{d_l} \big( \Delta^t_{k,l,i}\big)^2},
\qquad k\in\mathcal{C}_t,\; l\in L.
\label{eq_client-neuron-norm}
\end{equation}

Then, we sum up all local model magnitudes for $l$ as $||\Delta^t_{l}||_2$ according to Equation (\ref{eq_accumulated-neuron-magnitude}). The two neurons with the highest accumulated magnitudes, denoted as $f'$ and $g'$ ($f'<g'$), are recognized as source and target neurons. 

\begin{equation}
\lVert \Delta^t_{l}\rVert_2 \;=\; \sum_{k\in\mathcal{C}_t} \lVert \Delta^t_{k,l}\rVert_2,
\qquad l\in L.
\label{eq_accumulated-neuron-magnitude}
\end{equation}

Changes of parameters directly connected to the two neurons of each local model, $U^t=\{\Delta^t_{k,l}|c_k \in \mathbb{C}^t, l \in \{g', f'\}\}$, are extracted as critical features and fed into GMM to form two clusters: one good local model cluster and one bad local model cluster. Using the Uniform Manifold Approximation and Projection (UMAP) \cite{mcinnes2018umap} dimension reduction method, Figure \ref{fig_scatter} visualizes the comparison between malicious and benign updates in FL-RCC based on three different feature types. It shows that, compared to utilizing the parameters for the entire model or the output layer parameters, leveraging neuron-wise parameters is more effective in distinguishing poisoned models under TLFAs from benign ones, as two more obvious clusters are observed.

Compared to hard clustering methods, such as KMeans \cite{jebreel2024lfighter} and HDBSCAN \cite{LiuP:C:2025c}, GMM applies soft probabilistic clustering, avoiding rigid decision boundaries, thus making it suitable for heterogeneous environments (we consider this issue both in feature extraction to ensure detection accuracy). When the distinction between poisoned and benign models is blurry, GMM captures this uncertainty by modeling the underlying distribution of model parameters, enabling more flexible and calibrated detection. The denser cluster is identified as the bad one (as in \cite{jebreel2024lfighter} and as shown in Figure \ref{fig_scatter}), so its local models will be filtered out before global aggregation.  

Moreover, once source and target classes (the attack goal) are known via the described neuron-wise magnitude analysis, we can calculate SRec and ASR during the training to consistently monitor global model performance regarding the two classes. If the SRec value drops or the ASR value increases significantly compared to the values in the last round, i.e., value changes achieve the pre-defined thresholds $SRec^{thr}$ and $ASR^{thr}$, we discard this round's global model for robustness, as it may still be poisoned even after discarding the deemed adversarial local model contributions.    

\textbf{Malicious Client Exclusion:} To reduce the damage to the global model caused by data poisoning by malicious clients across FL rounds, we maintain a rating score $r_k \in [r^{min}, r^{max}]$ for each client $c_k$, which is updated in each round $t$ according to Equation (\ref{eq_rating}), where $\beta, \gamma \in (0, r^{max}]$ are both constants that control the corresponding rating reward and penalty steps, respectively.

\begin{equation}\label{eq_rating}
    r_k(t) = 
    \begin{cases}
    \displaystyle
    \min\bigl\{\,r_k(t-1) + \beta,\;r^{\mathrm{max}}\,\bigr\},     & \text{if $\omega^t_{k}$ benign},\\[1ex]
    \displaystyle
    \max\bigl\{\,r_k(t-1) - \gamma,\;r^{\mathrm{min}}\,\bigr\},    & \text{if $\omega^t_{k}$ poisoned}.
    \end{cases}
\end{equation}

The $r_k$ value decreases if the model parameters uploaded by $c_k$ are deemed adversarial in the current round, and increases if the detection result is deemed benign. The update rule accumulates per-client behavior over time so that transient or noisy detection errors do not immediately lead to exclusion. Once $r_k \leq r^{min}$, $c_k$ is identified as malicious with high confidence, thus excluded from the future client selection process, avoiding malicious clients consistently poisoning the global model training process. Compared to the count-based strategy \cite{LiuP:C:2025c}, the cumulative score reduces false positives caused by temporary conditions (e.g., small local data, stochastic training effects) while remaining sensitive to persistent, adversarial behavior.

\textbf{Complexity Analysis:} Assume the dimensionalities of the whole DNN model, the output layer of DNN, and one neuron in the output layer are $d_w$, $d_o$, and $d_e$, respectively. Note that $d_w\gg d_o\gg d_e$. The computation overhead of DEFEND in each round includes the following parts: 1) $\mathcal{O}\, (Md_o)$ to calculate the output layer changes of $M$ clients; 2) $\mathcal{O}\, (MEd_e)$ to compute the neuron-wise magnitudes of $E$ output neurons and $M$ clients; 3) $\mathcal{O}\, (E\, log\, E)$ to identify the source and target neurons from $E$ neurons; 4) $\mathcal{O}\, (Md_e)$ to cluster neuron-wise parameters of $M$ clients via GMM; and 5) $\mathcal{O}\, (M)$ to maintain blacklist and rating values for $M$ clients. Such that, the overall complexity of DEFEND is $\mathcal{O}\, (Md_o)$. Compared to other countermeasures based on the entire model, the output layer, or K-Means, e.g., Median ($\mathcal{O}\, (M\, log\, Md_w)$), TMean ($\mathcal{O}\, (M\, log\, Pd_w)$), Krum ($\mathcal{O}\, (M^2d_w)$), and FoolsGold ($\mathcal{O}\, (M^2d_o)$), in brief, DEFEND is computation-efficient.

\textbf{Practical Considerations:} As mentioned in the system model, we leverage standardized V2X security and privacy protocols for DEFEND, notably VPKI and pseudonyms, to ensure unlinkability, authenticity, and non-repudiation. Current countermeasures primarily focus on model-level misbehavior detection, without misbehaving client exclusion. With the mentioned V2X protocols, DEFEND can smoothly implement the client rating and exclusion strategy after the poisoned model detection strategy. Moreover, the neuron-wise magnitude analysis can identify the attack goal of TLFAs, even if adaptive adversaries change their goals during the training. DEFEND takes full advantage of such information for critical feature extraction and local model performance validation in each round, thus maximizing its mitigation effect.

\begin{figure}[tb]
\centerline{\includegraphics[width=0.48\textwidth]{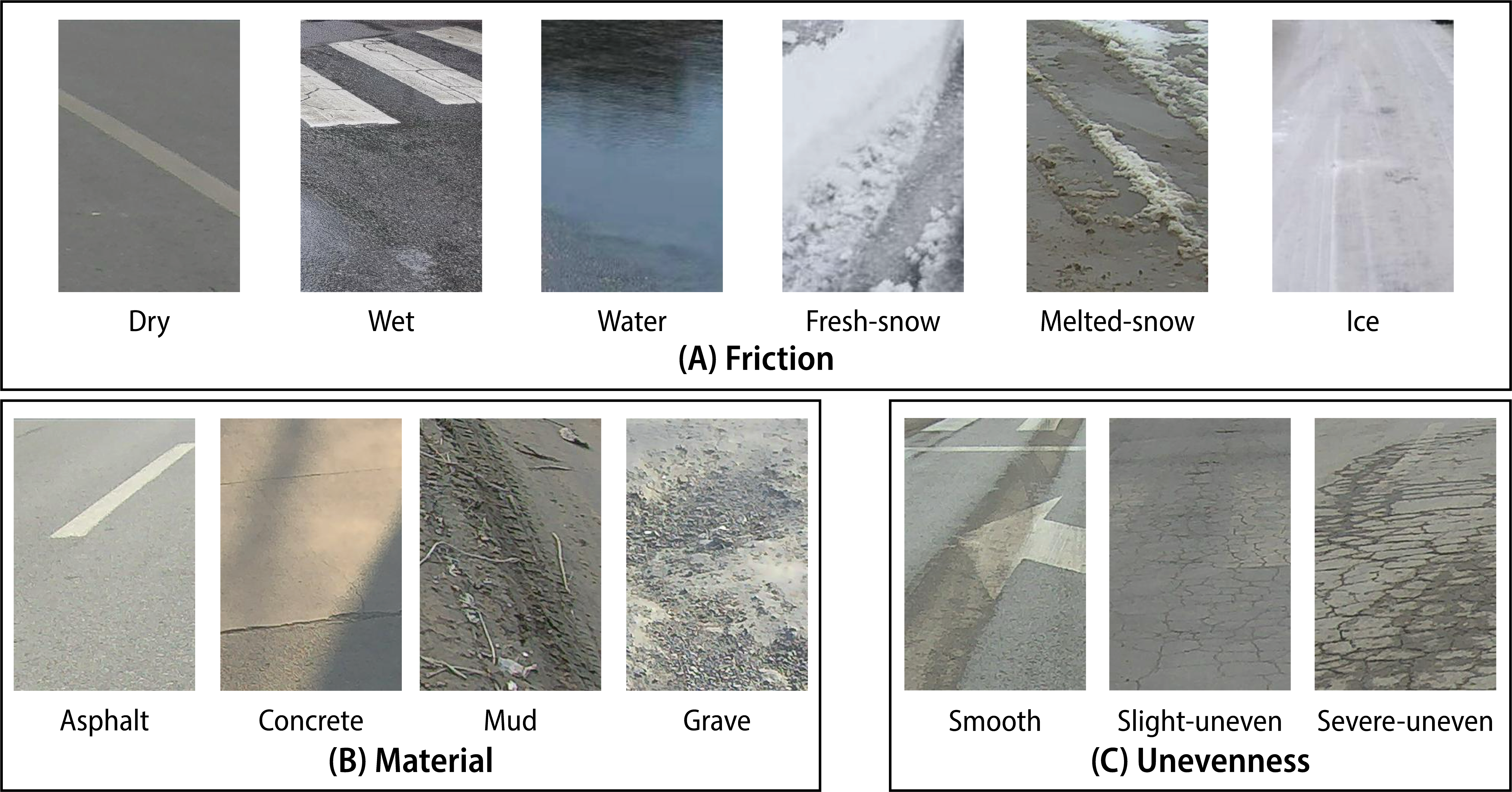}}
\vspace{-0.5em}
\caption{Image examples of the RSCD dataset.}
\vspace{-0.8em}
\Description{}
\label{fig_rscd}
\end{figure}

\begin{table}[tb]
\centering
\vspace{-0.5em}
\caption{Default model training configurations in this paper.}
\vspace{-1.2em}
\label{tab_config}
\centering
\begin{tabular}{p{0.213\textwidth}p{0.212\textwidth}}
\hline
\textbf{Term}   &\textbf{Value}  \\  \hline
Local epoch        & 3                \\
Global round       & 60                \\
Learning rate (lr) & 0.03              \\
Momentum for lr    & 0.5  \\
Optimizer          & SGD                \\
Batch size         & 64                 \\
Loss function      & Cross-Entropy \\
\hline
\end{tabular}
\end{table} 

\begin{table}[tb]
\centering
\vspace{-0.5em}
\caption{Practical model information of the Friction task.}
\vspace{-1.2em}
\label{tab_model}
\centering
\renewcommand{\arraystretch}{1.0} 
\resizebox{\columnwidth}{!}{%
\begin{tabular}{p{0.115\textwidth}p{0.085\textwidth}p{0.085\textwidth}p{0.1\textwidth}}
\hline
\textbf{Model}    &\textbf{Model} &\textbf{Inference} &\textbf{Memory}  \\
\textbf{Version}    &\textbf{Size (MB)} &\textbf{Time (ms)} &\textbf{Usage (MB)}  \\\hline
ResNet-18                    & 21.32          &2.18   &150.52   \\
EfficientNet-B1              & 12.44          &9.16   &237.30   \\
DeiT-Tiny                    & 10.54          &4.38   & 97.55   \\
\hline
\end{tabular}
}
\end{table}

\begin{table*}[tb]
  \scriptsize
  \centering
  \caption{The overall evaluation results within default configurations. All values are ratios in \%.}
  \label{tab_results}
  \renewcommand{\arraystretch}{1.0} 
  \begin{tabular}{C{1.90cm}C{1.51cm}||C{1.01cm}|C{1.01cm}|C{1.01cm}||C{1.01cm}|C{1.01cm}|C{1.01cm}||C{1.01cm}|C{1.01cm}|C{1.01cm}}
    \toprule
    \multirow{2.5}{*}{Model} & \multirow{2.5}{*}{Method} & \multicolumn{3}{c||}{RCC @ Friction} & \multicolumn{3}{c||}{RCC @ Material}  & \multicolumn{3}{c}{RCC @ Unevenness}\\
    \cmidrule(r){3-11}
    &  & GAcc $\uparrow$ & SRec $\uparrow$& ASR $\downarrow$ & GAcc $\uparrow$ & SRec $\uparrow$& ASR $\downarrow$ & GAcc $\uparrow$ & SRec $\uparrow$& ASR $\downarrow$ \\
    \midrule
    \multirow{10}{*}{ResNet-18 \cite{he2016deep}} & FedAvg-NA$^{\ddagger}$\cite{mcmahan2017communication} & $85.26$ & $72.28$ & $5.84$   & $80.36$ & $82.24$& $3.07$   & $74.65$ & $67.90$ & $9.93$ \\
    \cmidrule(r){2-11}
     & FedAvg \cite{mcmahan2017communication}    & $72.83$ & $44.88$ & $30.85$ & $70.17$ & $35.49$& $44.13$ & $69.14$ & $46.74$ & $50.34$ \\
     & Krum \cite{blanchard2017machine}      & $81.44$ & $48.89$ & $22.92$ & $58.87$ & $23.09$& $21.84$ & $48.54$ & $45.70$ & $53.38$ \\
     & TMean \cite{yin2018byzantine}     & $80.78$ & $45.84$ & $18.48$ & $55.75$ & $34.27$& $40.43$ & $52.65$ & $44.37$ & $\underline{25.23}$\\
     & Median \cite{yin2018byzantine}    & $81.56$ & $48.44$ & $22.52$ & $73.87$ & $46.85$& $25.07$ & $65.45$ & $47.33$ & $34.55$ \\   
     & FoolsGold \cite{fung2020limitations} & $81.64$ & $50.64$ & $15.56$ & $72.53$ & $42.53$& $35.20$ & $\underline{70.71}$ & $49.35$ & $26.33$ \\
     & FLAME \cite{nguyen2022flame} & $63.32$ & $51.20$ & $21.28$ & $52.51$ & $40.66$ & $45.20$ & $41.36$ & $16.29$ & $69.52$ \\
     & FLARE \cite{LiuP:C:2025c} & $\underline{82.80}^\dagger$ & $\underline{61.72}$ & $\underline{14.64}$ & $\underline{77.09}$ & $\underline{64.08}$ & $\underline{14.69}$ & $67.98$ & $\underline{55.13}$ & $28.13$ \\
     \rowcolor{lightgray}
     \cellcolor{white}& DEFEND (Ours)     & $\textbf{84.93}^\ast$ & $\textbf{74.20}$ & $\textbf{2.84}$ & $\textbf{78.61}$ & $\textbf{80.11}$& $\textbf{2.88}$ & $\textbf{73.71}$ & $\textbf{79.85}$ & $\textbf{7.32}$ \\
     \bottomrule
   \midrule

    \multirow{10}{*}{EfficientNet-B1 \cite{tan2019efficientnet}} & FedAvg-NA & $86.08$ & $75.68$ & $3.40$  & $80.15$ & $79.25$& $4.75$   & $74.48$ & $80.53$ & $8.08$ \\
    \cmidrule(r){2-11}
     & FedAvg     & $80.48$ & $38.12$ & $32.96$ & $74.63$ & $51.17$& $25.20$ & $65.48$ & $39.42$ & $29.42$ \\
     & Krum       & $64.49$ & $4.92$ & $69.32$ & $52.64$ & $42.88$& $20.13$ & $66.98$ & $43.67$ & $26.65$ \\
     & TMean      & $81.46$ & $46.52$ & $27.04$ & $69.30$ & $26.27$& $47.39$ & $66.86$ & $43.55$ & $26.78$\\
     & Median     & $82.54$ & $53.88$ & $22.16$ & $75.49$ & $54.59$ & $20.19$ & $66.63$ & $44.93$ & $28.02$ \\   
     & FoolsGold  & $\underline{83.64}$ & $59.80$ & $16.76$ & $\underline{77.84}$ & $61.15$& $17.84$ & $66.48$ & $38.32$ & $26.32$ \\
     & FLAME & $82.56$ & $52.36$ & $20.96$  & $59.19$ & $31.97$ & $14.32$ & $70.47$ & $50.25$ & $20.15$ \\
     & FLARE & $83.46$ & $\underline{62.04}$ & $\underline{16.08}$  & $77.56$ & $\underline{65.73}$ & $\underline{12.88}$ & $\underline{70.60}$ & $\underline{59.27}$ & $\underline{17.65}$ \\
     \rowcolor{lightgray}
     \cellcolor{white}& DEFEND (Ours)      & $\textbf{84.43}$ & $\textbf{80.40}$ & $\textbf{5.40}$ & $\textbf{77.87}$ & $\textbf{82.99}$& $\textbf{3.31}$ & $\textbf{72.87}$ & $\textbf{79.13}$ & $\textbf{7.98}$ \\
    \bottomrule
   \midrule

    \multirow{10}{*}{Deit-Tiny \cite{touvron2021training}} & FedAvg-NA & $86.54$ & $77.32$ & $3.44$  & $80.31$ & $80.61$& $3.79$   & $74.08$ & $78.80$ & $7.02$ \\
    \cmidrule(r){2-11}
     & FedAvg     & $77.89$ & $25.52$ & $48.40$ & $\underline{72.83}$ & $44.88$& $30.85$ & $60.93$ & $30.03$ & $45.47$ \\
     & Krum       & $67.15$ & $17.48$ & $38.04$ & $72.82$ & $50.48$& $31.81$ & $66.94$ & $40.73$ & $27.60$ \\
     & TMean      & $64.03$ & $34.88$ & $40.84$ & $47.37$ & $41.76$& $45.55$ & $64.93$ & $41.73$ & $37.50$ \\
     & Median    & $78.91$ & $33.64$ & $45.72$ & $51.81$ & $47.33$ & $41.28$ & $\textbf{77.01}$ & $\underline{55.72}$ & $\underline{14.65}$  \\   
     & FoolsGold  & $82.61$ & $53.60$ & $25.08$ & $\textbf{76.85}$ & $\underline{65.23}$& $\underline{14.75}$ & $68.59$ & $45.23$ & $17.33$ \\
     & FLAME & $57.54$ & $45.60$ & $25.84$  & $42.36$ & $43.57$ & $19.92$ & $60.34$ & $49.47$ & $33.63$ \\
     & FLARE & $\underline{83.02}$ & $\underline{58.84}$ & $\underline{17.80}$  & $72.19$ & $52.27$ & $29.09$ & $68.77$ & $53.70$ & $18.88$ \\
     \rowcolor{lightgray}
     \cellcolor{white}& DEFEND (Ours)   & $\textbf{84.58}$ & $\textbf{80.76}$ & $\textbf{4.84}$ & $65.55$ & $\textbf{82.11}$& $\textbf{4.75}$ & $\underline{69.14}$ & $\textbf{78.52}$ & $\textbf{7.63}$ \\
    \bottomrule
  
  \end{tabular}
\begin{tablenotes} 
	\footnotesize
	\item$^\ast$ \textbf{Bold numbers are the best performances in a group.}
	\item$^\dagger$ \underline{Numbers with underline are the second-best values in a group.} 
    \item$^{\ddagger}$ NA denotes No Attack. Others without this symbol are all under TLFAs.
\end{tablenotes}
\vspace{-1.0em}
\end{table*}

\section{Evaluation}\label{sec_eva}

\subsection{Experimental Setup}

We simulate 100 clients and one server in the PyTorch environment, using an NVIDIA A100 GPU (with 40GB memory) and an Icelake CPU (with 128GB memory) for the entire system. By default, 30 clients are malicious, and 20 clients are randomly selected as participants in a round, out of the total of 100 simulated clients. We evaluate each method's performance on three RCC tasks (Friction, Material, and Unevenness) based on the Road Surface Classification Dataset (RSCD)\footnote{\url{https://thu-rsxd.com/rscd/}} that comprises one million real‑world road images acquired from on-board cameras. We create non‑IID data partitioning for the 100 clients following the state of the art \cite{LiuP:C:2025c} from a Dirichlet($\alpha$) distribution with $\alpha=1.0$ by default.

For computational efficiency, all images are resized to $224\times224\times3$. Three task‑specific subsets include: 1) Friction: 58,800 training and 14,550 test samples across six classes (dry, wet, water, fresh‑snow, melted‑snow, and ice); 2) Material: 57,000 training and 15,000 test images spanning four surface types (asphalt, concrete, mud, and gravel); and 3) Unevenness: 57,542 training and 18,000 test images labeled by three degrees of roughness (smooth, slight‑uneven, and severe‑uneven). During training, each adversarial client applies TLFAs by relabeling its local samples as follows: water$\rightarrow$dry (Friction), gravel$\rightarrow$asphalt (Material), and severe-uneven$\rightarrow$smooth (Unevenness). Image examples of each subset are provided in Figure \ref{fig_rscd}. The global test sets remain untouched and serve solely for inference. Model training details, such as learning rate and batch size, are summarized in Table \ref{tab_config} as in \cite{LiuP:C:2025c}.

We adopt three lightweight (suitable for vehicular deployment) DNN models: ResNet‑18 \cite{he2016deep}, EfficientNet‑B1 \cite{tan2019efficientnet}, and DeiT‑Tiny \cite{touvron2021training}. The model size, inference time per image, and memory usage of each model completing the Friction task are summarized in Table \ref{tab_model}. We evaluate seven schemes in the experiment: FedAvg \cite{mcmahan2017communication} (not countering adversarial behavior but the basis for other baseline schemes), Krum \cite{blanchard2017machine}, TMean and Median \cite{yin2018byzantine}, FoolsGold \cite{fung2020limitations}, FLAME \cite{nguyen2022flame}, FLARE \cite{LiuP:C:2025c} and our proposed DEFEND ($SRec^{thr}=0.1$, $ASR^{thr}=0.1$, $r^{max}=1.00$, $r^{min}=0.00$, $r_k(0)=0.80$, $\beta=0.05$, and $\gamma=0.20$; these parameters are chosen empirically). All schemes share the same training configurations in Table \ref{tab_model} for fair comparison. Performance is quantified by three metrics: GAcc, SRec, and ASR.

\begin{figure*}[ht]
\centerline{\includegraphics[width=0.97\textwidth]{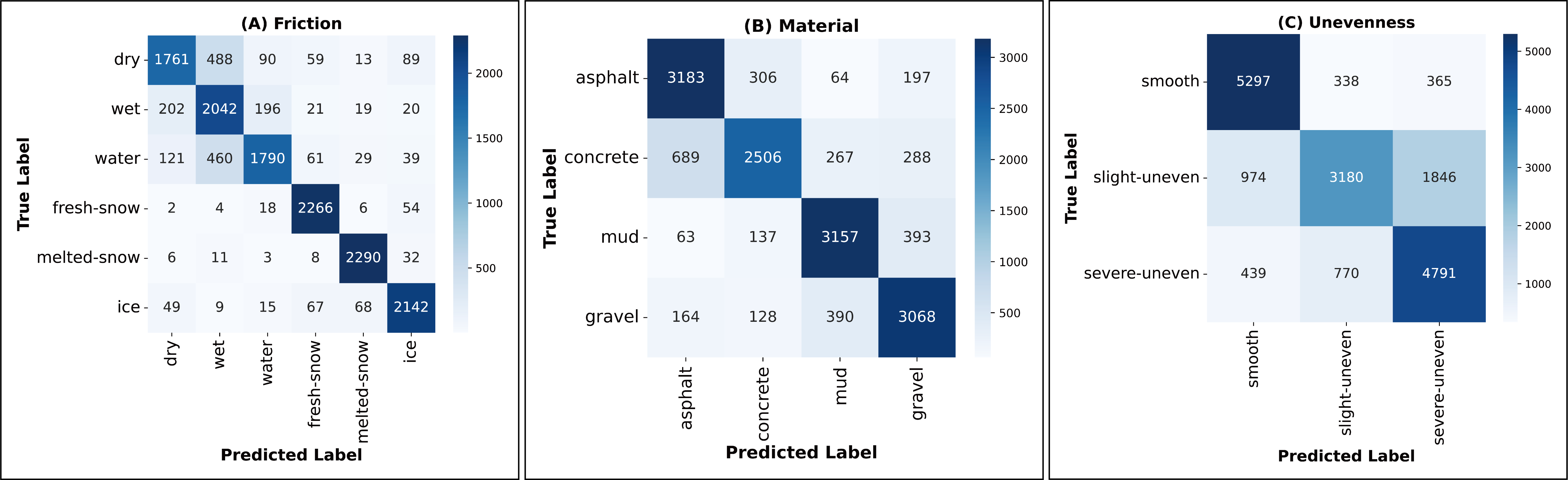}}
\vspace{-1.3em}
\caption{Confusion matrices of DEFEND with ResNet-18 in three RCC tasks.}
\Description{}
\label{fig_confu_resnet}
\end{figure*}

\begin{figure*}[ht]
\vspace{-1.0em}
\centerline{\includegraphics[width=0.97\textwidth]{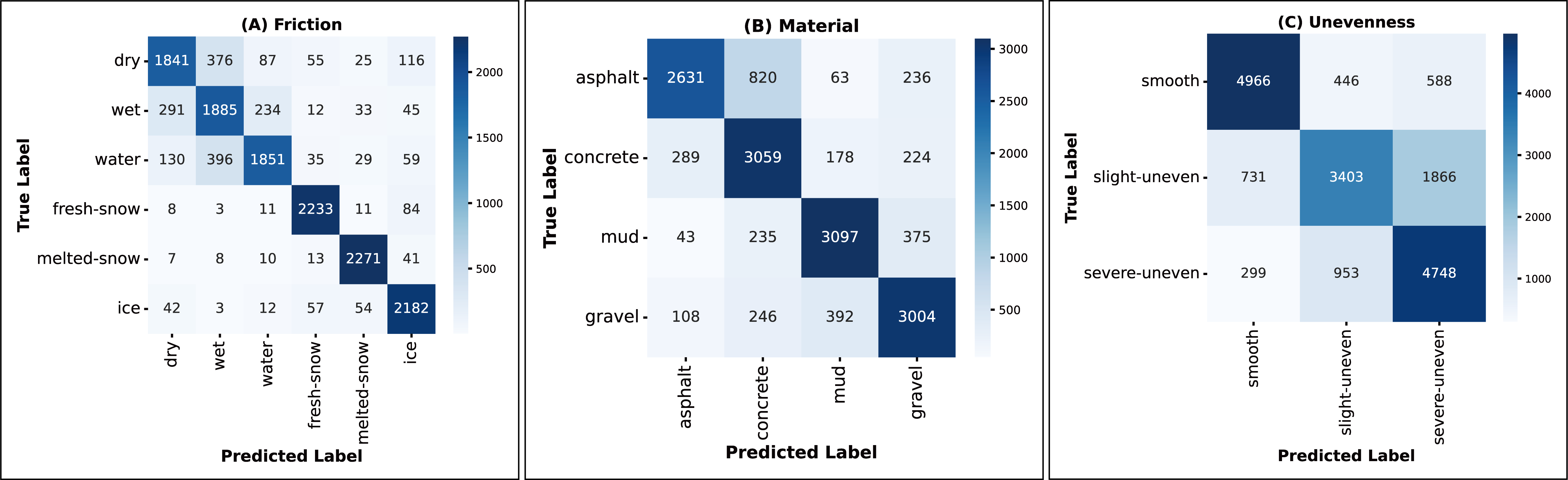}}
\vspace{-1.3em}
\caption{Confusion matrices of DEFEND with EfficientNet-B1 in three RCC tasks.}
\Description{}
\label{fig_confu}
\end{figure*}

\begin{figure*}[ht]
\vspace{-1.0em}
\centerline{\includegraphics[width=0.97\textwidth]{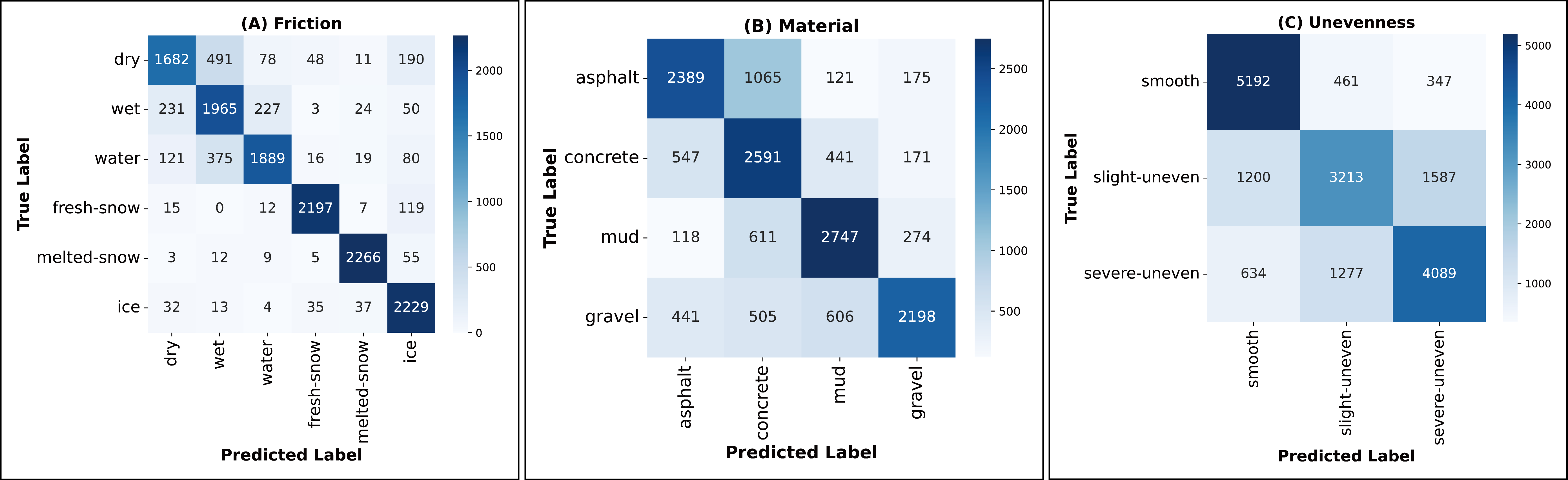}}
\vspace{-1.3em}
\caption{Confusion matrices of DEFEND with Deit-Tiny in three RCC tasks.}
\Description{}
\label{fig_confu_deit}
\end{figure*}

\begin{figure*}[ht]
\centerline{\includegraphics[width=0.97\textwidth]{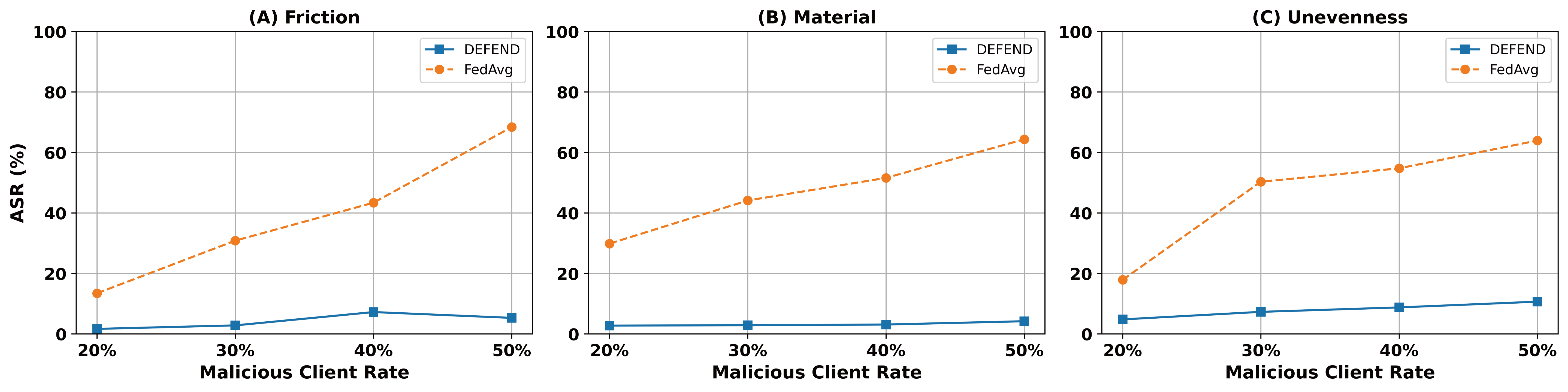}}
\vspace{-1.5em}
\caption{Impact of malicious client rates with ResNet-18 in three RCC tasks.}
\Description{}
\label{fig_poi_resnet}
\end{figure*}

\begin{figure*}[ht]
\vspace{-1.0em}
\centerline{\includegraphics[width=0.97\textwidth]{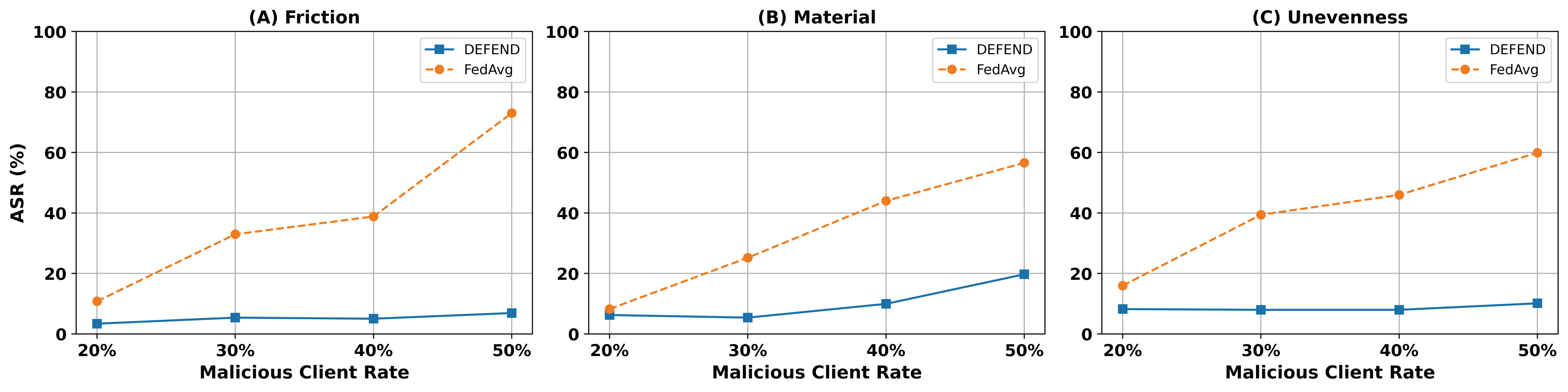}}
\vspace{-1.5em}
\caption{Impact of malicious client rates with EfficientNet-B1 in three RCC tasks.}
\Description{}
\label{fig_poi}
\end{figure*}

\begin{figure*}[ht]
\vspace{-1.0em}
\centerline{\includegraphics[width=0.97\textwidth]{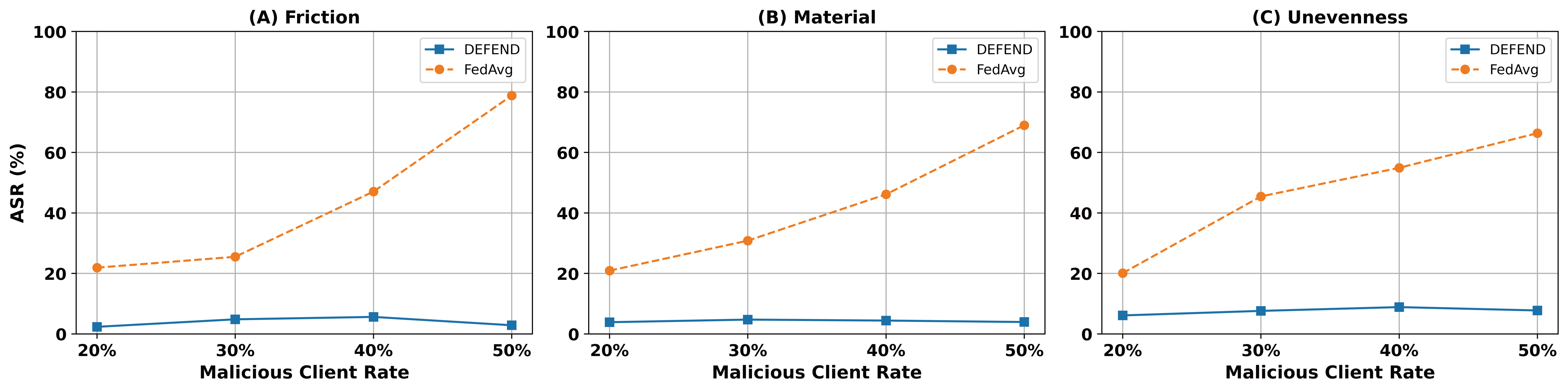}}
\vspace{-1.5em}
\caption{Impact of malicious client rates with Deit-Tiny in three RCC tasks.}
\Description{}
\label{fig_poi_deit}
\end{figure*}

\subsection{Results Analysis}\label{sec_result_anal}

\textbf{Baseline Results:} As per Table \ref{tab_results}, compared to FedAvg-NA (No Attack), FedAvg performance under TLFAs drops significantly, a 37.60\% average reduction in SRec and a 32.00\% average increase in ASR, indicating that FL-RCC is vulnerable to TLFAs. Although the six baseline countermeasures (Krum, TMean, Median, FoolsGold, FLAME, and FLARE) improve performance compared to FedAvg, there is still a huge performance gap to the FedAvg-NA performance. Take ResNet-18 with Unevenness as an example, the SRec values of the baseline countermeasures range from 16.29\% to 55.13\%, while the ASR values range from 25.32\% to 69.52\%. Even with the best of the baseline defenses, there is still significant risk: more than half of severe-uneven road conditions are recognized as smooth conditions, threatening safety. Specifically, the best baseline performance in each group of Table \ref{tab_results} (per model type and task type for fair comparison) is still worse than FedAvg-NA on average by 16.32\% and 11.01\% in terms of SRec and ASR, respectively. Such results indicate that the TLFAs mitigation of FL-RCC by existing countermeasures is limited, with stronger defense required before FL-RCC deployment.

\textbf{DEFEND Results:} In contrast, our method, DEFEND, remarkably boosts SRec and reduces ASR at the same time. On average, $2.81\%$, $24.99\%$, and $15.78\%$ improvement against the best baseline in each group are observed for DEFEND regarding GAcc, SRec, and ASR, respectively. The results show that DEFEND can accurately detect poisoned models and timely exclude malicious clients. Surprisingly, the average SRec and ASR values of DEFEND are $79.79\%$ and $5.22\%$, respectively, even slightly better than FedAvg-NA ($77.18\%$ and $5.48\%$, respectively), further indicating the effectiveness of DEFEND. The confusion matrices of DEFEND in three tasks are provided in Figure \ref{fig_confu_resnet} (with ResNet-18), Figure \ref{fig_confu} (with EfficientNet-B1), and Figure \ref{fig_confu_deit} (Deit-Tiny). These confusion matrices clearly show for all classes, including source and target classes per RCC task being the attack goals (water$\rightarrow$dry, gravel$\rightarrow$asphalt, and severe-uneven$\rightarrow$smooth), have high prediction performance (diagonal values). Such results further indicate that DEFEND can effectively prevent attackers from achieving goals that threaten transportation safety. Moreover, the nine confusion matrices show that DEFEND can work well for all three RCC tasks and three DNN models, indicating the stability and compatibility of DEFEND.

\textbf{Impact of Malicious Client Rate:} DEFEND is robust with ASR values always at a very low level as malicious client rate increases, for all three DNN models and three RCC tasks. As shown in Figure \ref{fig_poi_resnet} (with ResNet-18), Figure \ref{fig_poi} (with EfficientNet-B1), and Figure \ref{fig_poi_deit} (with Deit-Tiny), when malicious client rates gradually increase from 20\% to 50\% with a step of 10\%, the ASR for DEFEND does not significantly grow accordingly. Take EfficientNet-B1 for example: ASR ranges are [3.40\%, 6.92\%] for the Friction, [5.40\%, 19.73\%] for the Material, and [7.98\%, 10.13\%] for the Unevenness tasks. Even when half of clients are malicious, the model validation process in DEFEND, along with the preceding neuron-wise magnitude analysis, can still detect severely poisoned global models and discard them. However, the FedAvg ASR performance deteriorates seriously as the malicious client rate rises: 73.04\% (from 10.80\%) for the friction, 56.59\% (from 8.24\%) for the Material, and 59.93\% (from 15.97\%) for the Unevenness tasks. Increased malicious client rates significantly increase TLFA impact, while DEFEND can still work well even when the malicious client rate is very high, regardless of tasks and models.

\textbf{Observations on Model Type:} ResNet-18, as a lightweight Convolutional Neuron Network (CNN), suffers considerable degradation without defense mechanisms in place, while it benefits the most from DEFEND in terms of ASR. On average, EfficientNet-B1 delivers the strongest baseline performance in no-attack settings, but also experiences the steepest drops under attack. In contrast, the lightweight Transformer model, DeiT-Tiny, is highly vulnerable, with ASR frequently exceeding 40\% under FedAvg; while DEFEND markedly reduces ASR to match attack-free baselines, its performance remains less stable than CNN counterparts. These results show that CNNs offer increased robustness for FL-RCC, whereas Transformers require stronger defenses for comparable robustness.

\section{Conclusion}\label{sec_con}

This paper proposes a defensive mechanism, DEFEND, to secure FL-RCC systems against TLFAs. In each round, DEFEND detects poisoned models through neuron-wise magnitude analysis and GMM, and it validates global model performance after recognizing source and target classes. Moreover, based on model-level detection results, DEFEND adaptively rates clients and excludes those distrusted clients. Extensive evaluations involving various models, tasks, baselines, and metrics indicate the superiority of DEFEND over the state of the art: DEFEND under attack maintains the same model performance as in an attack-free scenario. Our future work shall extend DEFEND to autonomous driving tasks beyond RCC, e.g., behavioral intention prediction, explore machine unlearning and knowledge distillation to correct the already poisoned global model, and implement DEFEND in real vehicle environments, e.g., using NVIDIA Jetson-based edge devices.



\bibliographystyle{ACM-Reference-Format}
\bibliography{sample-base}

@String{Computing = "Computing" }

@String{Computer = "{IEEE} Computer" }

@String{Springer = "Springer-Verlag" }

@article{PapadimitratosH:J:2006,
	title        = {{Secure Data Communication in Mobile Ad Hoc Networks}},
	author       = {Papadimitratos, Panos and Haas, Z.J.},
	year         = {2006},
	_month        = {February},
	journal      = {IEEE Journal on Selected Areas in Communications},
	volume       = {24},
	number       = {2},
	pages        = {343--356},
	_doi          = {10.1109/JSAC.2005.861392},
	_issn         = {0733-8716},
	_keywords     = {ad hoc networks;data communication;fault tolerance;message authentication;mobile radio;routing protocols;telecommunication network topology;telecommunication security;MANET;SMT protocol;data communication;fault-tolerant communication;mobile ad hoc network;multihop wireless network;multipath routing overhead;network topology;secure message transmission;transmission failure detection;Data communication;Disruption tolerant networking;Fault tolerance;Mobile ad hoc networks;Mobile communication;Protocols;Spread spectrum communication;Surface-mount technology;Telecommunication network reliability;Wireless networks;Fault tolerance;mobile ad hoc network (MANET) security;multipath routing;network security;secure data transmission;secure message transmission;secure routing},
	_comment      = {DIVA},
	_webpdf       = {https://nss.proj.kth.se/publications/fulltext/secure-data-communication-manet.pdf},
	typ          = {J}
}

@ARTICLE{10480248,
  author={Zhou, Yijie and Cai, Likun and Cheng, Xianhui and Zhang, Qiming and Xue, Xiangyang and Ding, Wenchao and Pu, Jian},
  journal={IEEE Transactions on Intelligent Vehicles}, 
  title={OpenAnnotate2: Multi-Modal Auto-Annotating for Autonomous Driving}, 
  year={2024},
  volume={},
  number={},
  pages={1-13}}

@techreport{rescorla2018transport,
  title={The transport layer security (TLS) protocol version 1.3},
  author={Rescorla, Eric},
  year={2018}
}

@ARTICLE{KhodaeiP:J:2021b,
  author={Khodaei, Mohammad and Papadimitratos, Panos},
  journal={IEEE Transactions on Mobile Computing}, 
  title={Scalable \& Resilient Vehicle-Centric Certificate Revocation List Distribution in Vehicular Communication Systems}, 
  year={2021},
  volume={20},
  number={7},
  pages={2473-2489}}

@article{KhodaeiNP:J:2023,
  author={Khodaei, Mohammad and Noroozi, Hamid and Papadimitratos, Panos},
  journal={IEEE Transactions on Cloud Computing}, 
  title={SECMACE+: Upscaling Pseudonymous Authentication for Large Mobile Systems}, 
  year={2023},
  volume={11},
  number={3},
  pages={3009-3026}
  }

@article{KhodaeiJP:J:2018,
	author={Khodaei, Mohammad and Jin, Hongyu and Papadimitratos, Panagiotis},
  journal={IEEE Transactions on Intelligent Transportation Systems}, 
  title={SECMACE: Scalable and Robust Identity and Credential Management Infrastructure in Vehicular Communication Systems}, 
  year={2018},
  volume={19},
  number={5},
  pages={1430-1444}
}

@ARTICLE{10075082,
  author={IEEE Std 1609.2},
  journal={IEEE Std 1609.2-2022 (Revision of IEEE Std 1609.2-2016)}, 
  title={IEEE Standard for Wireless Access in Vehicular Environments--Security Services for Application and Management Messages}, 
  year={2023},
  volume={},
  number={},
  pages={1-349}}

@ARTICLE{4689252,
  author={Papadimitratos, Panagiotis and Buttyan, Levente and Holczer, Tamas and Schoch, Elmar and Freudiger, Julien and Raya, Maxim and Ma, Zhendong and Kargl, Frank and Kung, Antonio and Hubaux, Jean-Pierre},
  journal={IEEE Communications Magazine}, 
  title={Secure vehicular communication systems: design and architecture}, 
  year={2008},
  volume={46},
  number={11},
  pages={100-109}}

@article{varona2020deep,
  title={A Deep Learning Approach to Automatic Road Surface Monitoring and Pothole Detection},
  author={Varona, Braian and Monteserin, Ariel and Teyseyre, Alfredo},
  journal={Personal and Ubiquitous Computing},
  volume={24},
  number={4},
  pages={519--534},
  year={2020},
  publisher={Springer}
}

@ARTICLE{8896015,
  author={Basavaraju, Akanksh and Du, Jing and Zhou, Fujie and Ji, Jim},
  journal={IEEE Sensors Journal}, 
  title={A Machine Learning Approach to Road Surface Anomaly Assessment Using Smartphone Sensors}, 
  year={2020},
  volume={20},
  number={5},
  pages={2635-2647}}

@INPROCEEDINGS{8569396,
  author={Nolte, Marcus and Kister, Nikita and Maurer, Markus},
  booktitle={ITSC}, 
  title={Assessment of Deep Convolutional Neural Networks for Road Surface Classification}, 
  year={2018},
  volume={},
  number={},
  _pages={381-386},
  location={Maui, HI, USA}
}

@inproceedings{3670434,
author = {Lavaur, L\'{e}o and Busnel, Yann and Autrel, Fabien},
title = {Systematic Analysis of Label-flipping Attacks against Federated Learning in Collaborative Intrusion Detection Systems},
year = {2024},
_isbn = {9798400717185},
_publisher = {Association for Computing Machinery},
_address = {New York, NY, USA},
_url = {https://doi.org/10.1145/3664476.3670434},
_doi = {10.1145/3664476.3670434},
booktitle = {ARES},
_articleno = {63},
_pages={1-12},
_numpages = {12},
keywords = {backdoors, data-poisoning, federated learning, intrusion detection, label-flipping, quantitative assessment, systematic analysis},
location = {Vienna, Austria},
_series = {ARES '24}
}

@InProceedings{CONTRA,
author="Awan, Sana
and Luo, Bo
and Li, Fengjun",
_editor="Bertino, Elisa
and Shulman, Haya
and Waidner, Michael",
title="CONTRA: Defending Against Poisoning Attacks in Federated Learning",
booktitle="ESORICS",
year="2021",
_publisher="Springer International Publishing",
_address="Cham",
_pages="455-475",
location="Virtual",
isbn="978-3-030-88418-5"
}

@ARTICLE{10705359,
  author={Otoofi, Mohammad and Laine, Leo and Henderson, Leon and Midgley, William J. B. and Justham, Laura and Fleming, James},
  journal={IEEE Transactions on Intelligent Transportation Systems}, 
  title={FrictionSegNet: Simultaneous Semantic Segmentation and Friction Estimation Using Hierarchical Latent Variable Models}, 
  year={2024},
  volume={25},
  number={12},
  pages={19785-19795}}

@article{CHEN2025125978,
title = {{Enhancing Road Surface Recognition via Optimal Transport and Metric Learning in Task-Agnostic Intelligent Driving Environments}},
journal = {Expert Systems with Applications},
volume = {266},
pages = {125978},
year = {2025},
issn = {0957-4174},
_doi = {https://doi.org/10.1016/j.eswa.2024.125978},
_url = {https://www.sciencedirect.com/science/article/pii/S0957417424028458},
author = {Yuyi Chen and Shichun Yang and Rui Wang and Zhuoyang Li and Qiuyue Li and Zexiang Tong and Yaoguang Cao and Fan Zhou},
keywords = {Road surface recognition, Intelligent driving, Multi-task learning, Adversarial training, Metric learning}
}

@ARTICLE{10054157,
  author={Jiang, Yifeng and Zhang, Weiwen and Chen, Yanxi},
  journal={IEEE Transactions on Information Forensics and Security}, 
  title={Data Quality Detection Mechanism Against Label Flipping Attacks in Federated Learning}, 
  year={2023},
  volume={18},
  number={},
  pages={1625-1637}}

@ARTICLE{9641742,
  author={Yamany, Waleed and Moustafa, Nour and Turnbull, Benjamin},
  journal={IEEE Transactions on Intelligent Transportation Systems}, 
  title={OQFL: An Optimized Quantum-Based Federated Learning Framework for Defending Against Adversarial Attacks in Intelligent Transportation Systems}, 
  year={2023},
  volume={24},
  number={1},
  pages={893-903}}

@ARTICLE{10046398,
  author={Zhou, Hongliang and Zheng, Yifeng and Huang, Hejiao and Shu, Jiangang and Jia, Xiaohua},
  journal={IEEE Transactions on Intelligent Transportation Systems}, 
  title={Toward Robust Hierarchical Federated Learning in Internet of Vehicles}, 
  year={2023},
  volume={24},
  number={5},
  pages={5600-5614}}

@inproceedings{nguyen2022flame,
  title={FLAME: Taming Backdoors in Federated Learning},
  author={Nguyen, Thien Duc and Rieger, Phillip and De Viti, Roberta and Chen, Huili and Brandenburg, Bj{\"o}rn B and Yalame, Hossein and M{\"o}llering, Helen and Fereidooni, Hossein and Marchal, Samuel and Miettinen, Markus and others},
  booktitle={USENIX Security},
  _pages={1415-1432},
  location={Boston, MA, USA},
  year={2022}
}

@INPROCEEDINGS{10422129,
  author={Vondikakis, Ioannis V. and Panagiotopoulos, Ilias E. and Dimitrakopoulos, George J.},
  booktitle={ITSC}, 
  title={An Adaptive Federated Learning Framework for Intelligent Road Surface Classification}, 
  year={2023},
  volume={},
  number={},
  _pages={4121-4126},
  location={Bilbao, Spain},
  keywords={Federated learning;Roads;Image edge detection;Data models;Safety;Security;Surface treatment;neural network;machine learning;federated learning;road surface classification}}

@article{YUAN2021385,
title = {FedRD: Privacy-preserving Adaptive Federated Learning Framework for Intelligent Hazardous Road Damage Detection and Warning},
journal = {Future Generation Computer Systems},
volume = {125},
pages = {385-398},
year = {2021},
issn = {0167-739X},
_doi = {https://doi.org/10.1016/j.future.2021.06.035},
_url = {https://www.sciencedirect.com/science/article/pii/S0167739X21002302},
author = {Yachao Yuan and Yali Yuan and Thar Baker and Lutz Maria Kolbe and Dieter Hogrefe},
keywords = {Edge-cloud computing, Federated learning, Differential privacy, Hazardous road damage detection and warning, Traffic safety, Latency}
}

@article{MALIN2019181,
title = {{Accident Risk of Road and Weather Conditions on Different Road Types}},
journal = {Accident Analysis \& Prevention},
volume = {122},
pages = {181-188},
year = {2019},
issn = {0001-4575},
_doi = {https://doi.org/10.1016/j.aap.2018.10.014},
author = {Fanny Malin and Ilkka Norros and Satu Innamaa}
}

@inproceedings{tan2019efficientnet,
  title={EfficientNet: Rethinking Model Scaling for Convolutional Neural Networks},
  author={Tan, Mingxing and Le, Quoc},
  booktitle={ICML},
  _pages={6105--6114},
  year={2019},
  location = {Long Beach, CA, USA},
  _organization={PMLR}
}

@article{jebreel2024lfighter,
  title={LFighter: Defending against the Label-flipping Attack in Federated Learning},
  author={Jebreel, Najeeb Moharram and Domingo-Ferrer, Josep and S{\'a}nchez, David and Blanco-Justicia, Alberto},
  journal={Neural Networks},
  volume={170},
  pages={111--126},
  year={2024},
  publisher={Elsevier}
}

@ARTICLE{AFM3D,
  author={Liu, Sheng and You, Linlin and Zhu, Rui and Liu, Bing and Liu, Rui and Yu, Han and Yuen, Chau},
  journal={IEEE Transactions on Intelligent Transportation Systems}, 
  title={AFM3D: An Asynchronous Federated Meta-Learning Framework for Driver Distraction Detection}, 
  year={2024},
  volume={25},
  number={8},
  pages={9659-9674}}

@inproceedings{shejwalkar2021manipulating,
  title={Manipulating the Byzantine: Optimizing Model Poisoning Attacks and Defenses for Federated Learning},
  author={Shejwalkar, Virat and Houmansadr, Amir},
  booktitle={NDSS},
  _pages={1-18},
  location={Virtual},
  year={2021}
}

@inproceedings{touvron2021training,
  title={Training Data-efficient Image Transformers \& Distillation through Attention},
  author={Touvron, Hugo and Cord, Matthieu and Douze, Matthijs and Massa, Francisco and Sablayrolles, Alexandre and J{\'e}gou, Herv{\'e}},
  booktitle={ICML},
  _pages={10347-10357},
  year={2021},
  location={Virtual},
  _organization={PMLR}
}

@inproceedings{he2016deep,
  title={{Deep Residual Learning for Image Recognition}},
  author={He, Kaiming and Zhang, Xiangyu and Ren, Shaoqing and Sun, Jian},
  booktitle={CVPR},
  _pages={770-778},
  location={Las Vegas, NV, USA},
  year={2016}
}

@inproceedings{fereidooni2023freqfed,
  title={FreqFed: A Frequency Analysis-Based Approach for Mitigating Poisoning Attacks in Federated Learning},
  author={Fereidooni, Hossein and Pegoraro, Alessandro and Rieger, Phillip and Dmitrienko, Alexandra and Sadeghi, Ahmad-Reza},
  booktitle={NDSS},
  _pages={1-16},
  _organization={The Internet Society},
  location={San Diego, CA, USA},
  year={2024}
}

@inproceedings{rieger2022crowdguard,
  title={{CrowdGuard: Federated Backdoor Detection in Federated Learning}},
  author={Rieger, Phillip and Krau{\ss}, Torsten and Miettinen, Markus and Dmitrienko, Alexandra and Sadeghi, Ahmad-Reza},
  booktitle={NDSS},
  _pages={1-18},
  _organization={The Internet Society},
  location={San Diego, CA, USA},
  year={2024}
}

@article{sameera2024lfgurad,
  title={LFGurad: A Defense against Label Flipping Attack in Federated Learning for Vehicular Network},
  author={Sameera, KM and Vinod, P and KA, Rafidha Rehiman and Conti, Mauro},
  journal={Computer Networks},
  volume={254},
  pages={110768},
  year={2024},
  publisher={Elsevier}
}

@inproceedings{fung2020limitations,
  title={The Limitations of Federated Learning in Sybil Settings},
  author={Fung, Clement and Yoon, Chris JM and Beschastnikh, Ivan},
  booktitle={RAID},
  _pages={301-316},
  location={San Sebastian, Spain},
  year={2020}
}

@inproceedings{yin2018byzantine,
  title={Byzantine-robust Distributed Learning: Towards Optimal Statistical Rates},
  author={Yin, Dong and Chen, Yudong and Kannan, Ramchandran and Bartlett, Peter},
  booktitle={ICML},
  _pages={5650-5659},
  year={2018},
  location={Stockholm, Sweden},
  _organization={Pmlr}
}

@inproceedings{blanchard2017machine,
author = {Blanchard, Peva and El Mhamdi, El Mahdi and Guerraoui, Rachid and Stainer, Julien},
title = {Machine learning with Adversaries: Byzantine Tolerant Gradient Descent},
year = {2017},
isbn = {9781510860964},
_publisher = {Curran Associates Inc.},
booktitle = {NeurIPS},
_pages = {118–128},
location={Long Beach, CA, USA},
_numpages = {11},
}

@article{zhao2023comprehensive,
  title={A Comprehensive Implementation of Road Surface Classification for Vehicle Driving Assistance: Dataset, Models, and Deployment},
  author={Zhao, Tong and He, Junxiang and Lv, Jingcheng and Min, Delei and Wei, Yintao},
  journal={IEEE Transactions on Intelligent Transportation Systems},
  volume={24},
  number={8},
  pages={8361--8370},
  year={2023},
  publisher={IEEE}
}

@inproceedings{mcmahan2017communication,
  title={Communication-efficient Learning of Deep Networks from Decentralized Data},
  author={McMahan, Brendan and Moore, Eider and Ramage, Daniel and Hampson, Seth and y Arcas, Blaise Aguera},
  booktitle={AISTATS},
  _pages={1273-1282},
  year={2017},
  location={Fort Lauderdale, FL, USA},
  _organization={PMLR}
}

@inproceedings{tolpegin2020data,
  title={Data Poisoning Attacks against Federated Learning Systems},
  author={Tolpegin, Vale and Truex, Stacey and Gursoy, Mehmet Emre and Liu, Ling},
  booktitle={ESORICS},
  _pages={480-501},
  year={2020},
  location={Guildford, UK},
  _organization={Springer}
}

@ARTICLE{10606293,
  author={Vondikakis, Ioannis V. and Panagiotopoulos, Ilias E. and Dimitrakopoulos, George J.},
  journal={IEEE Open Journal of Intelligent Transportation Systems}, 
  title={FedRSC: A Federated Learning Analysis for Multi-Label Road Surface Classifications}, 
  year={2024},
  volume={5},
  number={},
  pages={433-444}}

@inproceedings{LiuP:C:2025c,
	title        = {{Safeguarding Federated Learning-based Road Condition Classification}},
	author       = {Liu, Sheng and Papadimitratos, Panos},
	year         = {2025},
	_month        = {July},
	booktitle    = {{IEEE CNS}},
	location      = {Avignon, France},
    _pages={1-9},
	webpdf       = {https://nss.proj.kth.se/publications/fulltext/2025_Safeguarding_FL_1571149969_final.pdf},
	typ          = {C}
}

@inproceedings{khuu2024data,
  title={Data Poisoning Detection in Federated Learning},
  author={Khuu, Denise-Phi and Sober, Michael and Kaaser, Dominik and Fischer, Mathias and Schulte, Stefan},
  booktitle={ACM SAC},
  _pages={1549--1558},
  location={Avila, Spain},
  year={2024}
}

@article{you2023federated,
  title={Federated and Asynchronized Learning for Autonomous and Intelligent Things},
  author={You, Linlin and Liu, Sheng and Zuo, Bingran and Yuen, Chau and Niyato, Dusit and Poor, H Vincent},
  journal={IEEE Network},
  volume={38},
  number={2},
  pages={286--293},
  year={2023},
  publisher={IEEE}
}

@article{mcinnes2018umap,
  title={UMAP: Uniform Manifold Approximation and Projection for Dimension Reduction},
  author={McInnes, Leland and Healy, John and Melville, James},
  journal={arXiv preprint arXiv:1802.03426},
  year={2020}
}

@inproceedings{10.1145/3488932.3517395,
author = {Wang, Ning and Xiao, Yang and Chen, Yimin and Hu, Yang and Lou, Wenjing and Hou, Y. Thomas},
title = {FLARE: Defending Federated Learning against Model Poisoning Attacks via Latent Space Representations},
year = {2022},
_isbn = {9781450391405},
_publisher = {Association for Computing Machinery},
_address = {New York, NY, USA},
_url = {https://doi.org/10.1145/3488932.3517395},
_doi = {10.1145/3488932.3517395},
booktitle = {ASIA CCS},
_pages = {946–958},
_numpages = {13},
location = {Nagasaki, Japan},
_series = {ASIACCS '22}
}

@inproceedings{li2024backdoorindicator,
  title={BackdoorIndicator: Leveraging OOD Data for Proactive Backdoor Detection in Federated Learning},
  author={Li, Songze and Dai, Yanbo},
  booktitle={USENIX Security},
  _pages={4193--4210},
  location={Philadelphia, PA, USA},
  year={2024}
}

@article{KhodaeiP:J:2015,
	title        = {{The Key to Intelligent Transportation: Identity and Credential Management in Vehicular Communication Systems}},
	author       = {Mohammad Khodaei and Panos Papadimitratos},
	year         = {2015},
	_month        = {December},
	journal      = {IEEE Vehicular Technology Magazine},
	volume       = {10},
	number       = {4},
	pages        = {63--69},
	_doi          = {10.1109/MVT.2015.2479367},
	_issn         = {1556-6072},
	_keywords     = {intelligent transportation systems;mobile radio;public key cryptography;telecommunication security;VC systems;VPKI;credential management;intelligent transportation system;standardization effort;user privacy;vehicular communication system;vehicular public-key infrastructure;Automation;Intelligent vehicles;Management;Network security;Privacy;Security;Vehicle ad hoc networks;Vehicular and wireless technologies;Vehicular automation},
	_publisherurl = {https://ieeexplore.ieee.org/document/7317862},
	_webpdf       = {https://nss.proj.kth.se/publications/fulltext/key-ITS-identity-credential-management.pdf},
	typ          = {J}
}

\appendix

\end{document}